\documentclass[aps,prd,twocolumn,groupedaddress,amssymb,eqsecnum,showpacs,psfig]{revtex4}
\usepackage{graphicx}
\usepackage{bm}
\usepackage{dcolumn}
\usepackage{amsmath}

\begin{document}

\newcommand{\newc}{\newcommand}

\newc{\be}{\begin{equation}}
\newc{\ee}{\end{equation}}
\newc{\ba}{\begin{eqnarray}}
\newc{\ea}{\end{eqnarray}}
\newc{\bea}{\begin{eqnarray*}}
\newc{\eea}{\end{eqnarray*}}
\newc{\D}{\partial}
\newc{\ie}{{\it i.e.} }
\newc{\eg}{{\it e.g.} }
\newc{\etc}{{\it etc.} }
\newc{\etal}{{\it et al.}}
\newcommand{\nn}{\nonumber}

\newc{\ra}{\rightarrow}
\newc{\lra}{\leftrightarrow}
\newc{\lsim}{\buildrel{<}\over{\sim}}
\newc{\gsim}{\buildrel{>}\over{\sim}}
\title{A comparison of cosmological models using recent supernova data}
\author{S. Nesseris and L. Perivolaropoulos}
\email{http://leandros.physics.uoi.gr} \affiliation{Department of
Physics, University of Ioannina, Greece}
\date{\today}

\begin{abstract}
We study the expansion history of the universe up to a redshift of
z=1.75 using the 194 recently published SnIa data by Tonry et. al.
and Barris et. al. In particular we find the best fit forms of
several cosmological models and $H(z)$ ansatze, determine the best
fit values of their parameters and rank them according to
increasing value of $\chi_{min}^2$ (the minimum value of $\chi^2$
for each $H(z)$ ansatz). We use a prior of $\Omega_{0m} = 0.3$ and
assume flat geometry of the universe. No prior assumptions are
made about validity of energy conditions. The fitted models are
fourteen and include SCDM, LCDM, dark energy with constant
equation of state parameter $w$ (quiessence), third order
polynomial for $H(1+z)$, Chaplygin gas, Cardassian model,
$w(z)=w_0 + w_1 z$, $w(z)=w_0 + z
 w_1/(1+z)$, an oscillating ansatz for $H(z)$ etc. All these models with
 the exception of SCDM are consistent with the present data.
 However, the quality of the fit differs significantly among them and so
 do the predicted forms of $w(z)$ and $H(z)$ at best
 fit.  The worst fit among the data-consistent models considered corresponds to the
 simplest model LCDM ($\chi_{min}^2 = 198.7$ for $\Omega_{0m} = 0.34$) while the best fit is
 achieved by the three parameter oscillating ansatz ($\chi_{min}^2 =
 194.1$). Most of the best fit ansatze have an equation of state parameter $w(z)$ that varies
 between $w(z) \simeq -1$ for $z<0.5$ to $w(z) > 0$ for $z>1$.
 This implies that the sign of the pressure of the dark energy may
 be alternating as the redshift increases. The goodness of fit of the
 oscillating $H(z)$ ansatz lends further support to this
 possibility.
\end{abstract}

\maketitle

\section{Introduction}
One of the fundamental goals of cosmology is the understanding of
the global history of the universe. Using objects of approximately
known absolute luminocity (standard candles) in the nearby
universe provides the current rate of expansion. Using more
distant standard candles like type Ia supernovae (SnIa) makes it
possible to start seeing the varied effects of the universe's
expansion history. Such cosmological observations have
indicated\cite{Riess:1998cb} that the universe undergoes
accelerated expansion during recent redshift times. This
accelerating expansion has been attributed to a dark energy
component with negative pressure which can induce repulsive
gravity and thus cause accelerated expansion. The simplest and
most obvious candidate for this dark energy
\cite{Sahni:2004ai} is the cosmological
constant\cite{Sahni:1999gb} with equation of state $w={p \over
\rho}=-1$.

The extremely fine tuned value of the cosmological constant
required to induce the observed accelerated expansion has led to a
variety of alternative models where the dark energy component
varies with time. Many of these models make use of a homogeneous,
time dependent minimally coupled scalar field $\phi$
(quintessence\cite{Peebles:1987ek,Sahni:1999qe}) whose dynamics is
determined by a specially designed potential $V(\phi)$ inducing
the appropriate time dependence of the field equation of state
$w(z) = {{p(\phi)} \over {\rho(\phi)}}$. Given the observed
$w(z)$, the quintessence potential can in principle be determined.
Other physically motivated models predicting late accelerated
expansion include modified 
gravity\cite{Perrotta:1999am,Torres:2002pe,Nojiri:2003ni},
Chaplygin gas\cite{Kamenshchik:2001cp}, Cardassian
cosmology\cite{Freese:2002sq}, theories with compactified extra
dimensions\cite{Perivolaropoulos:2002pn,Perivolaropoulos:2003we},
braneworld models\cite{Sahni:2002dx} etc. Such cosmological models
predict specific forms of the Hubble parameter $H(z)$ as a
function of redshift $z$ in terms of arbitrary parameters. These
parameters are determined by fitting to the observed luminocity
distance $d_L (z)$ using the
relation\cite{Starobinsky:1998fr,Huterer:2000mj,Chiba:1998de}
(valid in a flat universe) \be \label{hz1} H(z)= c [{d\over {dz}}
({{d_L(z)}\over {1+z}})]^{-1} \ee This is easily derived using the
relation between $d_L (z)$ and the comoving distance $r(z)$ (where
$z$ is the redshift of light emission) \be \label{dlz1} d_L (z) =
r(z) (1+z) \ee and the light ray geodesic equation in a flat
universe $c \; dt = a(z) \; dr(z)$ where $a(z)$ is the scale
factor.

Another similar approach towards determining the expansion history
$H(z)$ is to assume an arbitrary ansatz for $H(z)$ which is not
necessarily physically motivated (it is `model independent') but
is specially designed to give a good fit to the data for $d_L
(z)$. Given a particular cosmological model (or ansatz) for $H(z;
a_1, ... ,a_n)$ where $a_1, ...,a_n$ are model parameters, the
maximum likelihood technique can be used to determine the best fit
values of parameters (with $1\sigma - 2\sigma$ errors) as well as
the goodness of the fit of the ansatz to the data. This technique
can be summarized as follows: The observational data consist of
$N$ apparent magnitudes $m_i (z_i)$ and redshifts $z_i$ with their
corresponding errors $\delta m_i$ and $\delta z_i$. Each apparent
magnitude is related to the corresponding luminocity distance
$d_L$ of the SnIa by \be \label{mz1} m(z)=M + 5 \; log_{10} [{{d_L
(z)}\over {Mpc}}] + 25 \ee where $M$ is the absolute magnitude
which is assumed to be constant for standard candles like Type Ia
SnIa. From equations (\ref{hz1}) and (\ref{mz1}) it becomes clear
that the luminocity distance $d_L (z)$ is the `meeting point'
between the observed apparent magnitude $m(z)$ and the theoretical
prediction $H(z)$.

The apparent magnitude can also be expressed in terms of the
dimensionless `Hubble-constant free' luminocity distance $D_L$
defined by \be \label{dlz2} D_L (z) = {{H_0 d_L (z)}\over c} \ee
as \be \label{mz2} m(z)={\bar M}(M,H_0) + 5\; log_{10}(D_L (z))
\ee where ${\bar M}$ is the magnitude zero point offset and
depends on $M$ and $H_0$ as \be \label{bm1} {\bar M} = M + 5\;
log_{10} ({{c/H_0}\over {1Mpc}}) + 25 \ee The zero point offset is
an additional model independent parameter that needs to be fit
along with the model parameters $a_1, ...,a_n$. However, since
${\bar M}$ is model independent its value from a specific good fit
can be used directly to other fits of model parameters. Thus the
observed $m_i (z_i)$ can be translated to $D_{Li}^{obs} (z_i)$
using equation (\ref{mz2}) for the best fit value of ${\bar
M}_{obs}$ obtained from nearby SnIa. The theoretically predicted
value $D_L^{th} (z)$ in the context of a given model
$H(z;a_1,...,a_n)$ can be obtained by integrating the equation
(\ref{hz1}) as \be \label{dth1} D_L^{th} (z) = (1+z) \int_0^z dz'
\; {{H_0}\over {H(z';a_1,...a_n)}} \ee The best fit values for the
parameters $a_1,...,a_n$ are found by minimizing the quantity \be
\label{chi2def} \chi^2 (a_1,...,a_n)=\sum_{i=1}^N
{{(log_{10}D_L^{obs}(z_i)-log_{10}D_L^{th}(z_i))^2}\over
{(\sigma_{log_{10}D_L (z_i)})^2 + ({{\partial log_{10} D_L
(z_i)}\over {\partial z_i}} \sigma_{z_i})^2}} \ee where $\sigma_z$
is the $1\sigma$ redshift uncertainty of the data and
$\sigma_{log_{10}D_L (z_i)}$ is the corresponding $1\sigma$ error
of $log_{10}D^{obs}_L (z_i)$.

The probability distribution for the parameters $a_1,...,a_n$
is\cite{press92} \be \label{pdist} P(a_1,...,a_n) = {\cal N}
e^{-\chi^2 (a_1,...,a_n)/ 2} \ee where ${\cal N}$ is a
normalization constant. If prior information is known on some of
the parameters $a_1,...,a_n$ then we can either fix the known
parameters using the prior information or `marginalize' i.e.
average the probability distribution (\ref{pdist}) around the
known value of the parameters with an appropriate 'prior'
probability distribution. Here we use the former approach (fix the
parameters with prior information) for simplicity. This
simplification has negligible effect on our results as it can be
verified by comparing some of our results with corresponding
results in the literature where marginalization has been used
(e.g. Ref. \cite{Choudhury:2003tj} for LCDM).

It is straightforward to minimize $\chi^2 (a_1,...,a_n)$ using
numerical libraries like NAG \cite{minuit,NAG} (see also
\cite{press92}) or packages like Mathematica \cite{wolfram} to
find $\chi_{min}^2 ({\bar a}_1,...,{\bar a}_n)$\cite{npmath} where
$\chi_{min}^2$ is the minimum obtained for the best fit parameter
values ${\bar a}_1,...,{\bar a}_n$. If $\chi_{min}^2/(N-n) \lsim
1$ the fit is good and the data are consistent with the considered
model $H(z; a_1,...,a_n)$.

The variable $\chi_{min}^2$ is random in the sense that it depends
on the random data set used. Its probability distribution is a
$\chi^2$ distribution for $N-n$ degrees of freedom \cite{press92}.
This implies that $68\% $ of the random data sets will give a
$\chi^2$ such that \be \label{dchi1} \chi^2 (a_1,...,a_n) - \chi^2
({\bar a}_1,...,{\bar a}_n ) \leq \Delta \chi_{1\sigma}^2 (n) \ee
where $\Delta \chi_{1\sigma}^2 (n)$ is $1$ for $n=1$, $2.3$ for
$n=2$, $3.53$ for $n=3$ etc. Thus equation (\ref{dchi1}) defines
closed ellipsoidal surfaces around ${\bar a}_1,...,{\bar a}_n $ in
the $n$ dimensional parameter space. The corresponding $1\sigma $
range of the parameter $a_i$ is the range of $a_i$ for points
contained within this ellipsoidal surface. Similarly, it can be
shown that $95.4 \% $ of the random data sets will give a $\chi^2$
such that \be \label{dchi2} \chi^2 (a_1,...,a_n) - \chi^2 ({\bar
a}_1,...,{\bar a}_n ) \leq \Delta \chi_{2\sigma}^2 (n) \ee where
$\Delta \chi_{2\sigma}^2 (n)$ is $4.0$ for $n=1$, $6.17$ for
$n=2$, $8.02$ for $n=3$ etc. Thus equation (\ref{dchi2}) defines
the $2\sigma $ ellipsoidal surface in parameter space and
similarly for higher $\sigma$'s.

\section{Cosmic Expansion History}

We now apply the above described maximum likelihood method using a
recently published data set consisting of 194 ($N=194$) SnIa
\cite{Tonry:2003zg,Barris:2003dq}. This is a subset of the total
of 253 published SnIa sample obtained by imposing constraints $A_V
<0.5$ (excluding high extinction) and $z>0.01$ (reducing peculiar
velocity effects). Each data point at redshift $z$ includes the
logarithm of the Hubble-free luminocity distance $log_{10}(c
D_L^{obs} (z))$ and the corresponding error $\sigma_{log_{10}D_L
(z)}$. A table of the data we used can be downloaded in electronic
form \cite{npmath}. These Hubble-free luminocity distances are
obtained assuming a best fit value for the zero point magnitude
offset ${\bar M}$ \cite{Choudhury:2003tj}. We adopt this same
value for ${\bar M}$ and choose not to treat ${\bar M}$ as an
additional free parameter to fit (and marginalize) along with the
parameters of each theoretical model studied. In the {\it
Appendix} we demonstrate that marginalization over ${\bar M}$
would have negligible effect $(O(1\%))$ on our results. Also
comparison of our results for LCDM and quiessence
($w(z)=constant$) with the corresponding results of Ref.
\cite{Choudhury:2003tj} where marginalization of ${\bar M}$ was
implemented indicates that our simplified approach has negligible
effect on the obtained results. This same conclusion has also been
reached in Ref. \cite{DiPietro:2002cz} 
and it's origin is demonstrated in the {\it Appendix}.

In the construction of $\chi^2$ using equation (\ref{chi2def}) we
have used a value of $\sigma_z$ corresponding to uncertainties due
to peculiar velocities with $\Delta v =\Delta(cz)=500km/sec$ which
implies $\sigma_z=\Delta z= (500km/sec)/c $. The minimization of
(\ref{chi2def}) was implemented for each theoretical model using a
simple Mathematica code which can be downloaded along with the
table of the data set used\cite{npmath} (or can be sent by e-mail
upon request).

We now proceed to apply likelihood testing to various theoretical
models. Each model is defined by its predicted Hubble-parameter
$H(z)$. For example for LCDM we have \be \label{lcdm}
H^2(z;\Omega_{0m}) = ({{{\dot a}}\over a})^2 = H_0^2 [\Omega_{0m}
(1+z)^3 + (1- \Omega_{0m})] \ee and there is a single parameter
$\Omega_{0m}$ to be fit from the data. The simplest model to
consider however is SCDM defined by  \be \label{SCDM} H^2(z)=H_0^2
(1+z)^3 \ee with no free parameters. Using equation (\ref{SCDM})
in (\ref{dth1}) we calculate $D_L^{th} (z)$. We may then find
$\chi^2$ using equation (\ref{chi2def}) and minimize to find
$\chi_{min}^2$. In the SCDM case there are no free parameters to
vary and no minimization is needed. We thus find
$\chi^2=\chi_{min}^2=431.4$ which implies $\chi_{min}^2 /dof =2.2$
($dof$=degrees of freedom). Since this value $\chi_{min}^2 /dof $
is significantly larger than $1$ we conclude that SCDM does not
provide a good fit to the SnIa data.

The next simplest model consistent with the flatness indicated by
WMAP\cite{deBernardis:2001xk} is LCDM defined by equation
(\ref{lcdm}). It is straightforward to evaluate $D_L^{th}(z)$
numerically (using equation (\ref{lcdm}) in (\ref{dth1})) and use
it to evaluate $\chi^2 (\Omega_{0m})$ from equation
(\ref{chi2def}). A minimization of this expression leads
to\cite{npmath} \be \label{chiminLCDM} \chi_{min}^2 = \chi^2
(\Omega_{0m}=0.34)=198.74 \ee which implies $\chi^2/dof = 1.03 $
($dof=194-1=193$). This model is clearly consistent with the data
since $\chi^2/dof \simeq 1$. The $1\sigma$ errors on the predicted
value of $\Omega_{0m} =0.34$ are found by solving the equation \be
\label{dchi3} \chi^2 (\Omega_{m1\sigma})-\chi_{min}^2 = \Delta
\chi_{1\sigma}^2 (n=1)=1 \ee which leads to \be \label{omer}
\Omega_{0m}=0.34 \pm 0.032 \ee This result is identical with the
result of Ref.\cite{Choudhury:2003tj} even though our $1\sigma$
errors are slightly smaller. We note here for comparison with the
models discussed below (where the prior $\Omega_{0m}=0.3$ is used)
that $\chi^2 (\Omega_{0m}=0.3)=200.3 $.

In Fig. 1 we show a comparison of the observed 194 SnIa Hubble
free luminocity distances along with the theoretically predicted
curves in the context of SCDM (continuous line) and LCDM (dashed
line). In this case it is even visually obvious that LCDM provides
a good fit to the data contrary to the case of SCDM.
\begin{figure}
\centering
\includegraphics[bb=60 100 380 700,width=8cm,height=10cm,angle=-90]{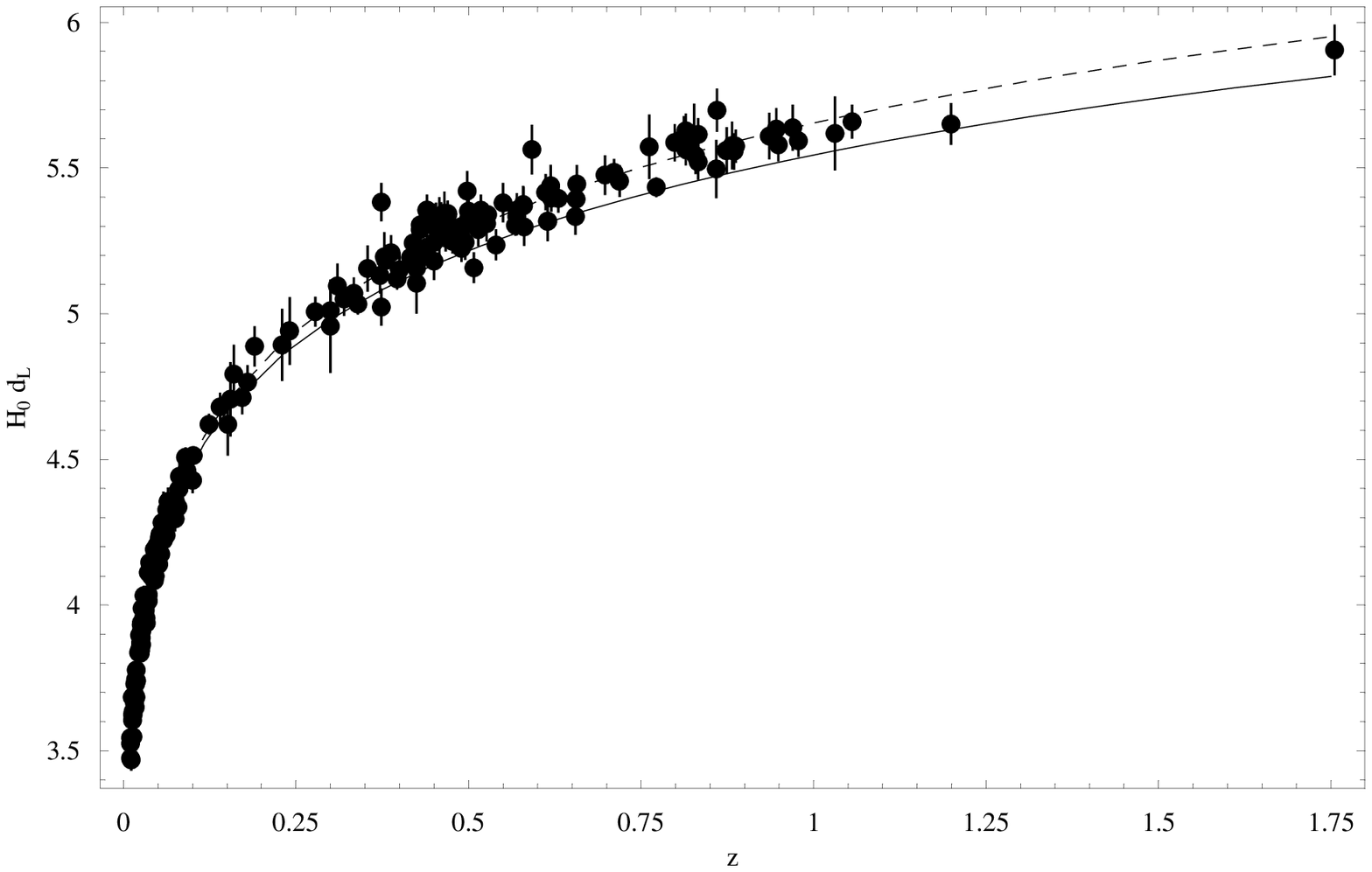}
\caption{The observed 194 SnIa Hubble free luminocity distances
along with the theoretically predicted curves in the context of
SCDM (continuous line) and LCDM (dashed line).} \label{fig1}
\end{figure}
This visual distinction is not possible when comparing the other
data-compatible models discussed below with LCDM. We thus do not
attempt to include the theoretical curves corresponding to other
models on the same plot.

We now consider other more general models and ansatze which
however reduce for certain parameter values to LCDM. If these
parameter values give a $\chi^2 (LCDM)$ that is beyond the
$2\sigma $ level away from the minimum $\chi_{min}^2 $ then we
would conclude that LCDM is disfavored compared to the better fit
model. Even if we just find models with $\chi_{min}^2 <
\chi_{min}^2 (LCDM)=198.74 $ but within $1\sigma$ we still have
useful information since these models are more probable than LCDM.

We start with a simple generalization of LCDM: We replace the
cosmological constant energy density by a dark energy with
constant equation of state parameter. This ansatz has been called
`quiessence' in the literature \cite{Alam:2003sc}. The form of
$H(z)$ is \be \label{hzw0} H^2(z;\Omega_{0m},w)=H_0^2 [
(\Omega_{0m} (1+z)^3 + (1-\Omega_{0m})(1+z)^{3(1+w)}] \ee This
ansatz has two free parameters $\Omega_{0m}$ and $w$. We use prior
information from large scale structure ($\Omega_{0m}h = 0.2\pm
0.03 $ \cite{2dF} with $h=0.72\pm0.08$ \cite{HST}) to fix
$\Omega_{0m} =0.3 $ in this and in all subsequent ansatze. We thus
evaluate $\chi^2 (w)$ and minimize with respect to $w$. We find
\be \label{cmw} \chi_{min}^2 = \chi^2 (w=-0.93)=199.3 \ee
Including the $1\sigma $ errors we have \be \label{wbf} w=-0.93
\pm 0.08 \ee For $\Omega_{0m}=0.34$ we find\cite{npmath} \be
\label{cmw1} \chi_{min}^2 = \chi^2 (w=-1.01)=198.69 \ee which is
identical with the corresponding result of
Ref.\cite{Choudhury:2003tj}. Thus, the minimization of this
generalized ansatz gives a best fit that is indistinguishable at
the $1\sigma$ level from LCDM. This means either that LCDM is
truly the best fit model or that we have not chosen a general
enough ansatz to see a better fit.

A further generalized ansatz involves the combination of
cosmological constant with quiessence (quiessence-$\Lambda$
ansatz).

\begin{figure}
\centering
\includegraphics[bb=35 100 450 730,width=8cm,height=8cm,angle=-90]{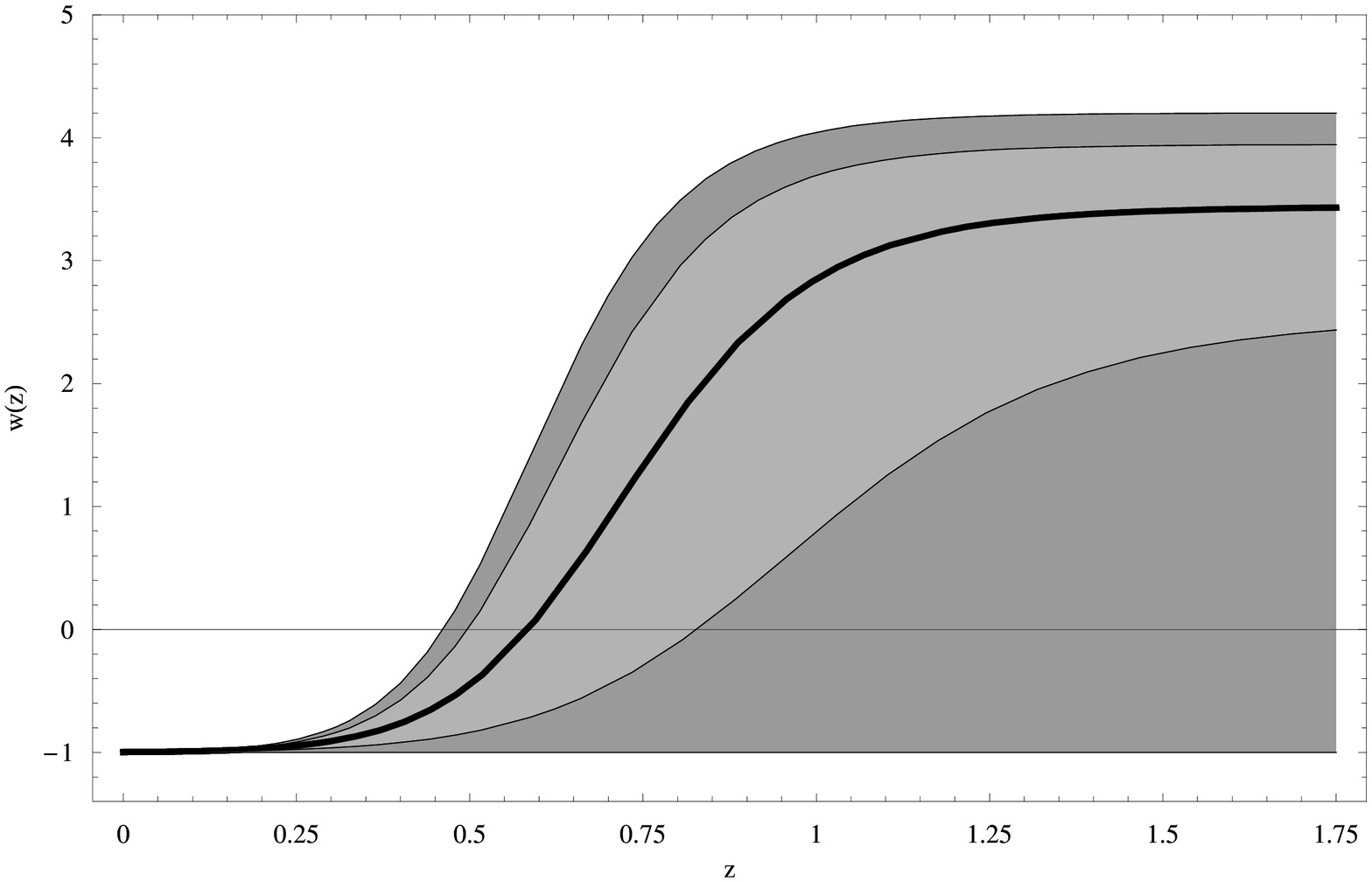}
\caption{The redshift dependence of the equation of state
parameter $w(z)$ for the $q-\Lambda$ ansatze. The thick curve is
the best fit and the light (dark) shaded regions represent the
$1\sigma$ ($2\sigma$) error regions.} \label{fig2}
\end{figure}

The form of $H(z)$ in this case is \ba \label{hzao} &
H^2(z;a_1,w_1) = H_0^2 [\Omega_{0m} (1+z)^3 + a_1 (1+z)^{3(1+w_1)}
+& \nn \\ & (1-\Omega_{0m} - a_1)]&  \ea Setting $\Omega_{0m} =
0.3$ and minimizing $\chi^2 (a_1,w_1)$ with respect to $w_1$,
$a_1$ we find \be \label{xmwa} \chi_{min}^2=\chi^2 
(w_1=3.44,\; a_1= (5\;10^{-4}) = 195.6 \ee Including the error
bars we have \be w_1=3.44^{+0.44}_{-0.78},\; a_1\simeq (5\
^{+0.4}_{-0.9})\;10^{-4} \ee Clearly the fit is better compared to
LCDM but the $\chi^2 (LCDM)=\chi^2 (a_1 = 0)=200.3$ corresponding
to LCDM with $\Omega_{0m}=0.3$ differs by less than $\Delta
\chi^2_{2\sigma}(n=2)=6.17 $ from $\chi_{min}^2$. Therefore, LCDM
is consistent at the $2\sigma$ level (but not at the $1\sigma$)
with the best fit of this ansatz. Nevertheless, given that this
fit is better it is interesting to compare the dark energy
properties corresponding to this ansatz at best fit with those of
LCDM. These properties are well described by the effective
equation of state parameter $w(z)={{p(z)}\over {\rho (z)}}$ which
in general (and in this case) depends on the redshift $z$. We can
express $w(z)$ in terms of $H(z)$, ${{dH}\over {dz}}$ and
$\Omega_{0m}$ using the Friedman equations \be \label{fr1} H^2
={{{\dot a}^2}\over a^2}={{8\pi G}\over 3} (\rho_m + \rho_{DE})
\ee and \be \label{fr2} q\equiv -{{{\ddot a}}\over {a H^2}} =
{{4\pi G}\over {3 H^2}} [\rho_m + (\rho_{DE} + 3 p_{DE})] \ee
where $q$ is the deceleration parameter and we have defined as
dark energy any other homogeneous and isotropic source of gravity
apart from matter. Using (\ref{fr1}) and (\ref{fr2}) we find \be
\label{pp} p_{DE}={{H^2}\over {4\pi G}} (q-{1\over 2}) \ee Using
(\ref{fr1}) and (\ref{pp}) we find\cite{Saini:1999ba} \be
\label{wz2} w(z)={{p_{DE}(z)}\over {\rho_{DE}(z)}}={{2 q(z)
-1}\over {3 (1-\Omega_m (z))}} \ee where \be \label{omz2} \Omega_m
(z) = {{8\pi G \rho_m (z)} \over {3 H^2 (z)}}= \Omega_{0m} (1+z)^3
{{H_0^2}\over {H^2}} \ee Using now the definitions of $q$ and H it
is easy to show that \be \label{qh} q=-1+(1+z) {{d\ln H}\over
{dz}} \ee Thus substituting  (\ref{qh}) in (\ref{wz2}) we have \be
\label{wz3} w(z)={{p_{DE}(z)}\over {\rho_{DE}(z)}}={{{2\over 3}
(1+z) {{d \ln H}\over {dz}}-1} \over {1-({{H_0}\over H})^2
\Omega_{0m} (1+z)^3}} \ee In the case of generalized Friedman
equations valid in modified gravity models, equation (\ref{wz3})
can still be useful in characterizing the expansion history but it
should not be interpreted as a property of an energy substance.
Using the best fit form of the quiessence-$\Lambda$ (q-$\Lambda$)
ansatz in (\ref{wz3}) we find the predicted form of $w(z)$ which
is plotted in Fig. 2 along with the $1\sigma$ and $2\sigma$ error
regions obtained by maximal variation of the parameters $a_1$ and
$w_1$ within the $1\sigma$ and $2\sigma$ error contours of
$\chi^2$ as described in the previous section. This form of $w(z)$
(without error regions) along with the corresponding forms
predicted by the other ansatze discussed below, is also shown in
Fig. 3. Clearly $w(z)$ differs significantly from the LCDM
prediction of $w=-1$ at redshifts $z>0.5$. In particular we find
$w(z)\simeq -1 $ for $z<0.5$ while $w(z) \simeq 3 $ for $z\gsim
1$. Thus, this ansatz gives us a hint for the `metamorphosis' of
dark energy from antigravity ($w=-1$) at low redshifts to
`hypergravity' ($w\simeq 3$) at high redshifts. Clearly this
`metamorphosis' (if true) can not persist to arbitrarily high
redshifts due to constraints coming from large scale structure and
nucleosynthesis. Thus, it is either not realized in nature and we
have $w(z) \leq 0$ at all redshifts or it is part of an
oscillating behavior of the dark energy equation of state
parameter. This later possibility could also help resolve the
coincidence problem\cite{Dodelson:2001fq,Griest:2002cu} and is a
prediction\cite{Perivolaropoulos:2002pn} (see also \cite{GZ,GZ1})
of many
models\cite{Arkani-Hamed:1999gq,Antoniadis:1990ew,Randall:1999vf}
with stabilized modulus\cite{Goldberger:1999uk,Csaki:1999mp} of
extra dimensions.

In addition to $w(z)$ we also plot the reduced form of $H(z)$
compared to LCDM defined as \be \label{hrz} H_r^2 (z) = {{H^2 (z)
- H_{LCDM}^2 (z)}\over {H_0^2}} \ee where $H_{LCDM}^2 \equiv 0.3
\; (1+z)^2 + 0.7 $. The reduced best fit $H_r^2 (z)$ for the
q-$\Lambda$ ansatz is shown in Fig. 4 along with the best fits
$H_r (z)$ functions corresponding to some of the other ansatze
discussed below. Even though both ansatze (q-$\Lambda$ and LCDM)
are consistent with the data and with each other at the $2\sigma$
level the predicted forms of $H_r (z)$ and $w(z)$ differ
significantly at $z>0.5$.

In order to shed more light to the dark energy `metamorphosis'
puzzle we now consider more general forms of $H(z)$ ansatze. For
each ansatz we identify the predicted form of $H(z)$ and the
parameters requiring fitting. Then we evaluate and minimize
$\chi^2$ with respect to these parameters setting
$\Omega_{0m}=0.3$. Finally, we identify the best fit parameter
values and the corresponding $\chi_{min}^2$ (see Table 1), plot
the corresponding $w(z)$ and $H(z)$ (see Figs. 3 and 4) and
classify the models according to the goodness of fit. This
classification will lead to some interesting conclusions about the
generic properties of dark energy.  The $1\sigma$ and $2\sigma$
regions of the $w(z)$ and $H(z)$ curves in the context of a
particular ansatz are not particularly useful since other ansatze
with equally good fits can give $w(z)$ and $H(z)$ best fits that
are well outside the $2\sigma$ regions of the initial ansatz
especially in regions with $z > 1$ where just a few data points
are available. Thus, in order to simplify the plots of Figs 3 and
4 and avoid confusion we only show the best fit curves without the
corresponding $1\sigma$ and $2\sigma$ regions.

We now briefly describe each one of the ansatze considered:
\begin{enumerate}
\item {\it Cubic Polynomial in (1+z) (P3):} \ba \label{p3} &&H^2
(z)=H_0^2 [\Omega_{0m} (1+z)^3 + a_3 (1+z)^3 + a_2 (1+z)^2 \nn
\\
&&+ a_1 (1+z) + (1-a_1-a_2 -a_3 - \Omega_{0m} )] \ea where $a_1,
\; a_2, \; a_3 $ are unknown parameters to be fit. The only priors
used in this (and the other ansatze discussed here) are
$\Omega_{0m}=0.3$ and flatness. In contrast to Ref.
\cite{Alam:2003fg} we have not fixed $a_3=0$ since large scale
structure data do not exclude a non-clustering form of matter with
properties similar to those of hot dark matter with very large
free streaming (or Jeans) length. This model has a slightly worse
fit ($\chi_{min}^2 = 196.6$) compared to q-$\Lambda$
($\chi_{min}^2 = 195.6$) but it also corresponds to $w(z)\simeq
-1$ at $z\lsim 0.5$ and $w(z)>0$ at $z \gsim 1$. Its properties at
best fit are shown in Figs. 3 and 4. 
The values of the best fit parameters are $a_1=-2.55\pm 0.12$,
$a_2=0.50 \ ^{+0.07}_{-0.05}$ and $a_3=0.36\pm 0.06 $. Compared to
the quadratic ansatz of Ref. \cite{Alam:2003fg} this cubic ansatz
has slightly better $\chi^2_{min}$ and $w(z)$ is larger than the
corresponding $w(z)$ of the quadratic ansatz at $z>1$ shown in
Ref. \cite{Alam:2003fg} and in Fig. 3.
 \item {\it
Linder\cite{Linder:2002et} ansatz $w(z)=w_0 + {{w_1 z}\over
{1+z}}$ (LA)}: \ba  \label{la} &H^2 (z)=H_0^2 [\Omega_{0m} (1+z)^3
+&  \nn \\ & (1-\Omega_{0m}) (1+z)^{3(1+w_0+w_1)} \; e^{3 w_1
({1\over {1+z}} -1)}]& \ea It can easily be verified using
(\ref{wz3}) that the ansatz (\ref{la}) leads to a $w(z)$ of the
form \be \label{law} w(z)=w_0 + {{w_1 z}\over {1+z}} \ee This
ansatz was suggested by Linder\cite{Linder:2002et} and it is
designed to interpolate between two values of $w$: $w(z)=w_0$ at
$z\simeq 0$ and $w(z)\simeq w_1$ at $z >> 1$. It does not give a
very good fit (compared to other two parameter ansatze) to the
data ($\chi_{min}^2 = 197.3$) even though it is still better than
LCDM. The reason is that it only allows a slow change of $w(z)$
with redshift between $w_0$ and $w_1$ while the data seem to
require a more `abrupt' change from $w_0$ to $w_1$. The best fit
values of these parameters are 
$w_0 = -1.29\pm 0.10 $ and $w_1 =2.84\pm 0.05$.
 \item {\it
Linear Ansatz $w(z) = w_0 + w_1 z$: } \ba \label{lin} && H^2
(z)=H_0^2 [\Omega_{0m} (1+z)^3 + \nn \\ &&(1-\Omega_{0m})
(1+z)^{3(1+w_0-w_1)} \; e^{3 w_1 z}] \ea This ansatz  was
suggested in Refs \cite{Astier:2000as,Weller:2001gf,Maor:2001ku}
in a more general power series form. It can give very good fits
for low redshift data ($z\lsim 0.5$) but it can not fit well the
general form of $w(z)$ at $z\gsim 1$. At best fit it gives
$\chi_{min}^2 = 196.6$ which is about average compared to the
other ansatze. The best fit parameter values are 
$w_0 = -1.25\pm0.09$ and $w_1 = 1.97 \ ^{+0.08}_{-0.01}$ which
leads to $w(z)> 0.5$ for $z\gsim 1 $.

\begin{figure}
\centering
\includegraphics[bb=10 100 350 670,width=8cm,height=10cm,angle=-90]{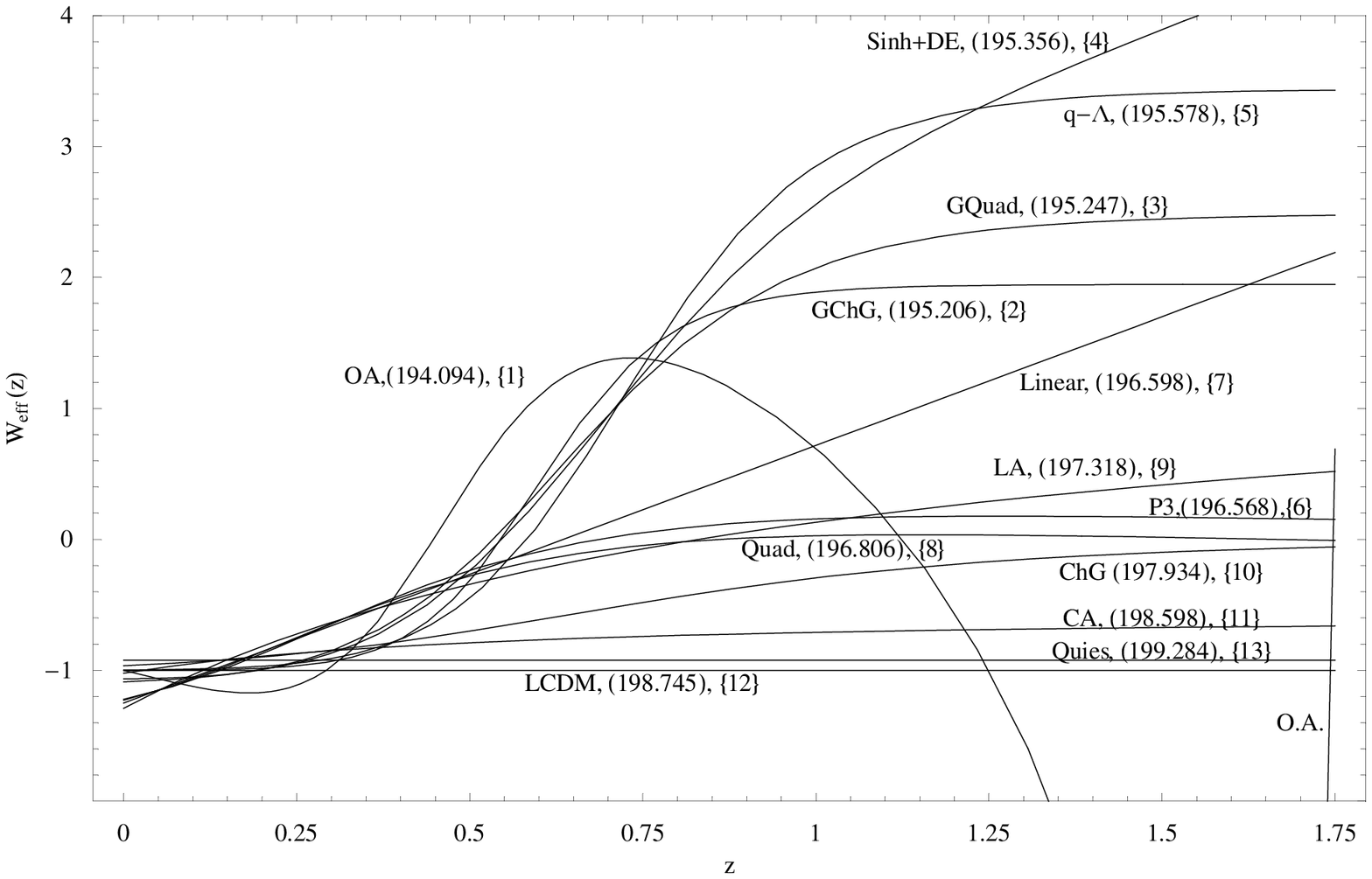}
\caption{The redshift dependence of the equation of state
parameter for the cosmological ansatze of Table 1. The numbers in
the parentheses indicate the value of $\chi^2_{min}$ for each
ansatz and its rank according to increasing value of
$\chi^2_{min}$. The prior $\Omega_{0m}=0.3$ has been used in all
cases except LCDM where the best fit value $\Omega_{0m}=0.34$ was
used giving $\chi_{min}^2 =198.745$.} \label{fig3}
\end{figure}

 \item
{\it Chaplygin gas (CG) and Generalized Chaplygin gas 
(GCG)\cite{Kamenshchik:2001cp,Bilic:2001cg,Fabris:2002vu,Alcaniz:2002yt,Avelino:2003cf,Amendola:2003bz,
Bento:2002ps,Bento:2003dj,Bento:2003we,Silva:2003bs,Bento:2002yx,Bertolami:2004ic,Gorini:2004by}:}
\ba \label{gcg} & H^2 (z)=H_0^2 [\Omega_{0m} (1+z)^3 + & \nn \\
&(1-\Omega_{0m}) \sqrt{A + (1-A) (1+z)^\alpha }]& \ea where the
above form of $H(z)$ is a generalization of the usual Chaplygin
gas ansatz which is obtained for $\alpha = 6$
\cite{Fabris:2002vu}. For $\alpha = 6$ the equation of state of
Chaplygin gas dark energy is \be \label{cges} p_c = - {A \over
{\rho_c}} \ee From the form of equation (\ref{gcg}) it is clear
that the Chaplygin gas behaves like pressureless dust at high
redshifts and like a cosmological constant at $z\simeq 0$. The
sound velocity of Chaplygin gas grows rapidly and approaches the
velocity of light at late redshifts ($v_s = \sqrt{{{dp_c} \over
{d\rho_c}}} = {{\sqrt{A}} \over {\rho_c}} \sim t^2 $). Thus
inhomogeneities of Chaplygin gas do not grow (since the Jeans
length approaches the horizon) and no constraints can be imposed
from large scale structure observations. The physical motivation
of the Chaplygin gas equation of state (\ref{cges}) comes from
string theories\cite{Ogawa:2000gj}. In fact, considering a $d$
brane in a $d+2$ dimensional space-time, the introduction of light
cone variables in the resulting Nambu-Goto action leads to the
action of a Newtonian fluid with equation of state (\ref{cges}).

The limitation of the Chaplygin gas ansatz is that it constrains
$w(z)$ to $w(z) <0$ at all finite redshifts. Thus its goodness of
fit is below average ($\chi_{min}^2 = 197.9$ for 
$A=0.96\pm0.03$). This limitation does not exist for the
generalized Chaplygin gas ansatz (arbitrary $\alpha$) which gives
a much better fit ($\chi_{min}^2 = 195.2$ for 
$A=0.9998 \ ^{+0.0001}_{-0.0004}$,\;$\alpha=17.68 \
^{+0.02}_{-0.04}$). This fit gives $w(z) \simeq 1.9> 0$ for
$z\gsim 1$ (see Fig. 3) as do all the ansatze with above average
goodness of fit.
 \item {\it Generalized Cardassian Ansatz
(CA):} \be \label{ca} H^2 (z)=H_0^2 [\Omega_{0m} (1+z)^3 + (1-
\Omega_{0m}) f_X (z)] \ee where \be \label{caf} f_X (z) =
{{\Omega_{0m}} \over {1-\Omega_{0m}}} (1+z)^3
[(1+{{\Omega_{0m}^{-q} - 1} \over {(1+z)^{3(1-n)q}}})^{1/q} - 1]
\ee This model \cite{Wang:2003cs,Zhu:2003sq} emerges from a
generalization of the Friedman equations and predicts accelerated
expansion at recent times without any dark energy. In this model
the universe is flat and consists only of matter and radiation.
Here we follow Ref. \cite{Wang:2003cs} and consider the
generalization (\ref{caf}) of the original Cardassian ansatz of
Ref. \cite{Freese:2002sq}. The original ansatz is obtained from
equation (\ref{caf}) by setting $q=1$ and is equivalent to
quiessence ($w=const.$). The generalized Cardassian ansatz has
been fitted to SnIa data in Ref. \cite{Wang:2003cs} using a much
smaller SnIa dataset with redshifts $z<1$. We find (see Table 1)
that this ansatz gives a relatively poor fit to the data
($\chi_{min}^2=198.6$ for 
$q=0.025\ ^{+0.008}_{-0.010}$, $n=-23\ ^{+8}_{-9}$) below the
average goodness of fit for the models considered and only
slightly better compared to LCDM with $\Omega_{0m} =0.3$
($\chi_{min}^2 = 200.4$). The predicted $w(z)$ increases with $z$
but remains negative (see Fig. 3).
 \item {\it Generalized Quardatic Ansatz (GQuad):} \ba \label{gq} & H^2 (z)=H_0^2
[\Omega_{0m} (1+z)^3  +  a_1 (1+z)^{3(1+w_1)} + & \nn \\ & a_2
(1+z)^{3(1+w_2)} + (1-a_1-a_2- \Omega_{0m} )]& \ea This is another
generalization of the quadratic polynomial fit for $H(z)$ of Ref.
\cite{Alam:2003fg}. Here we do not add an arbitrary cubic term.
Instead we allow the exponents of the two monomials to vary and
minimize with respect to the four parameters $a_1$, $a_2$, $w_1$
and $w_2$ instead of two parameters $a_1$ and $a_2$ with fixed
$w_1 = -2/3$ and $w_2 = -1/3$ for the quadratic model. The fit of
the generalized ansatz is  better ($\chi_{min}^2=195.2$ 
for $a_1=0.57\ ^{+0.03}_{-0.02}$, $a_2=0.003\
^{+0.0003}_{-0.0002}$, $w_1=-1.13\pm0.24$ and $w_2=2.49\pm0.02$)
than the quadratic ansatz ($\chi_{min}^2=196.8$ for 
$a_1=-4.05\ ^{+1.16}_{-1.27}$ and $a_2=1.79\ ^{+0.79}_{-0.63}$)
and the best fit form of $w(z)$ differs significantly from the
corresponding quadratic best fit particularly for $z\gsim 1$. In
particular we find that $w(z) \simeq -1$ for $z\lsim 0.4$ and
$w(z) \gsim 2$ for $z \gsim 1$. For comparison the quadratic
ansatz of Ref. \cite{Alam:2003fg} predicts $w(z) \simeq 0$ for
$z\gsim 1$ with very small $2\sigma$ errors at $z>1$. This
disagreement of our generalized ansatz with the quadratic ansatz
at $z>1$ despite the small $2\sigma $ error regions is another
indication of the limited usefulness of ploting $1\sigma$ and  $2
\sigma $ error regions of $w(z)$ in the context of a particular
ansatz. These regions can be easily violated in the context of
another ansatz with better or similar fit.

A variant of this ansatz is one where one of the two arbitrary
power law terms is replaced by an exponentially increasing term.
The corresponding ansatz (Sinh+DE) is \ba \label{sinh} & H^2
(z)=H_0^2 [\Omega_{0m} (1+z)^3  +  a_1 (1+z)^{3(1+w_1)} +& \nn \\
& a_2 \sinh(w_2 z) + (1-a_1 -\Omega_{0m} )]&  \ea The results for
this ansatz are almost identical to those of equation (\ref{gq})
(see Table 1 and Fig. 3). \item {\it Oscillating Ansatz (OA):} \ba
\label{oa} & H^2 (z)=H_0^2 [\Omega_{0m} (1+z)^3  +  a_1 \cos (a_2
z^2 + a_3) +& \nn
\\ & (1-a_1 \cos(a_3) - \Omega_{0m} )]& \ea This is our best fit
ansatz. It gives a better fit to the data than any of the other
ansatze ($\chi_{min}^2 = 194.1$ for 
$a_1 = -3.36\ ^{+0.93}_{-0.76}$, $a_2 = 2.12\ ^{+0.93}_{-0.76}$
and $a_3 = -0.06 \pi\pm0.01\pi $). The behavior of $w(z)$ however
for $z\gsim 1$ is qualitatively different compared to the other
ansatze (see Fig. 3). For $z\lsim 0.3$ we find $w(z) \simeq -1 $.
For $0.5 \lsim z \lsim 1.2$ we find $w(z)>0$ with a maximum
$w(z\simeq 0.75) \simeq 1.5$. At $z\gsim 1.2$, $w(z)$ becomes
negative and continues oscillating around $w\simeq 0$ with large
amplitude. This redshift range however includes only one data
point at $z=1.75$ which can not constrain the behavior of $w(z)$
and $H(z)$ in any statistically significant way.
\end{enumerate}

\begin{widetext}
\begin{center}
{\bf \large Table 1}

\begin{tabular}{|c|c|c|c|c|}
\hline {\bf Model}        &{\bf $H(z)$, ($\Omega_{0m}=0.3$) }    &  $\chi_{min}^2$ & {\bf Best Fit Parameters} \\
\hline OA & $H^2 (z)=H_0^2 [\Omega_{0m} (1+z)^3  +  a_1 \cos (a_2
z^2 + a_3)$ && $a_1=-3.36\ ^{+0.93}_{-0.76} ,\; a_2=2.12\ ^{+0.93}_{-0.76}$ \\ (1) & $+(1-a_1 \cos(a_3) - \Omega_{0m} )]$ & $194.1$ &$a_3=-0.06\pi\pm0.01\pi$ \\
\hline GCG (2) & $H^2 (z)=H_0^2 [\Omega_{0m} (1+z)^3
+(1-\Omega_{0m})\sqrt{A + (1-A) (1+z)^\alpha }]$ &$195.2$ &
$A=0.9998\ ^{+0.0001}_{-0.0004}, \; \alpha=17.68\
^{+0.02}_{-0.04}$ \\
\hline GQ  & $H^2 (z)=H_0^2 [\Omega_{0m} (1+z)^3  +  a_1
(1+z)^{3(1+w_1)} + a_2 (1+z)^{3(1+w_2)}$ &&$w_1=-1.13\pm 0.24, \; w_2=2.49\pm0.02$ \\ (3) & $+ (1-a_1-a_2- \Omega_{0m} )]$ & $195.2$ & $a_1=0.57\ ^{+0.03}_{-0.02}, \; a_2=0.003_{-0.0002}^{+0.0003}$ \\
\hline Sinh+DE & $H^2 (z)=H_0^2 [\Omega_{0m} (1+z)^3
+a_1(1+z)^{3(1+w_1)}+ a_2\sinh(w_2z)$ &&$w_1=-0.81\ ^{+0.1}_{-0.3}, \; w_2=5.90\ ^{+0.93}_{-3.13}$ \\ (4) & $+1-\Omega_{Om}-a_1]$ & $195.4$ &$a_1=-0.50\ ^{+0.1}_{-0.3} ,\; a_2=0.026\ ^{+0.005}_{-0.015}$ \\
\hline q-$\Lambda$  (5) &  $ H^2(z) = H_0^2 [\Omega_{0m} (1+z)^3 +
a_1 (1+z)^{3(1+w_1)} + (1-\Omega_{0m} - a_1) ]$    &$195.6$
&$w_1=3.44\ ^{+0.44}_{-0.78},\; a_1\simeq 5\;10^{-4}\
^{+0.4\;10^{-4}}_{-0.9\;10^{-4}}$       \\
\hline P3 & $ H^2 (z)=H_0^2 [\Omega_{0m} (1+z)^3 + a_3 (1+z)^3 +
a_2 (1+z)^2+$ &&$a_1=-2.55\pm 0.12$  \\ (6) & $a_1 (1+z) + (1-a_1-a_2 -a_3 - \Omega_{0m} )] $    &$196.6$ &$a_2=0.50\ ^{+0.07}_{-0.05}, \; a_3=0.36\pm 0.06 $  \\
\hline Linear & $H^2 (z)=H_0^2 [\Omega_{0m} (1+z)^3 +
(1-\Omega_{0m})$ && \\ (7) & $(1+z)^{3(1+w_0-w_1)}e^{3 w_1 z}]$ &
$196.6$ &$w_0=-1.25\pm0.09, \; w_1=1.97\ ^{+0.08}_{-0.01} $\\
\hline Quad  & $H^2 (z)=H_0^2 [\Omega_{0m} (1+z)^3  +  a_1
(1+z) + a_2 (1+z)^2$ && \\(8) & $ + (1-a_1-a_2- \Omega_{0m} )]$ & $196.8$ & $a_1=-4.05\ ^{+1.16}_{-1.27}, \; a_2=1.79\ ^{+0.79}_{-0.63}$ \\
\hline LA & $H^2 (z)=H_0^2 [\Omega_{0m} (1+z)^3 + (1-\Omega_{0m})
(1+z)^{3(1+w_0+w_1)}$ && \\(9) & $e^{3 w_1 ({1\over{1+z}
}-1)}] $ & $197.3$ &$w_0=-1.29\pm 0.10, \; w_1=2.84\pm 0.05 $\\
\hline CG (10)& $H^2 (z)=H_0^2 [\Omega_{0m} (1+z)^3 +(1-\Omega_{0m})\sqrt{A + (1-A) (1+z)^6 }]$ &$197.9$ & $A=0.96\pm0.03$\\
\hline CA (11)& $H^2 (z)=H_0^2 [\Omega_{0m} (1+z)^3 + (1-
\Omega_{0m})f_X (z)]$ &$198.6$ &  $q=0.03\ ^{+0.008}_{-0.010}, \; n=-23\ ^{+8}_{-9}$ \\
\hline LCDM (12)& $ H^2(z)=H_0^2 [\Omega_{0m}(1+z)^3 + (1- \Omega_{0m})]  $    &$198.7$ &$\Omega_{0m}=0.34\pm0.03$       \\
\hline QUIES (13)& $ H^2(z)=H_0^2 [\Omega_{0m}(1+z)^3 +
(1-\Omega_{0m}) 
(1+z)^{3(1+w)} ] $    &$199.3$   &$w=-0.93\pm0.08$\\
\hline SCDM & $H^2(z)=H_0^2 (1+z)^3 $ & $431.4$ &$-$       \\
\hline
\end{tabular}
\vspace{-0.2cm}
\end{center}
\end{widetext}

The quality of the fit for the oscillating ansatz combined with
indications from the better fits of other ansatze that indicate
$w(z)> 0$ for $z\gsim 1$ supports the idea that some type of
oscillation probably takes place for $w(z)$.

\begin{figure}
\centering
\includegraphics[bb=20 100 370 700,width=8cm,height=10cm,angle=-90]{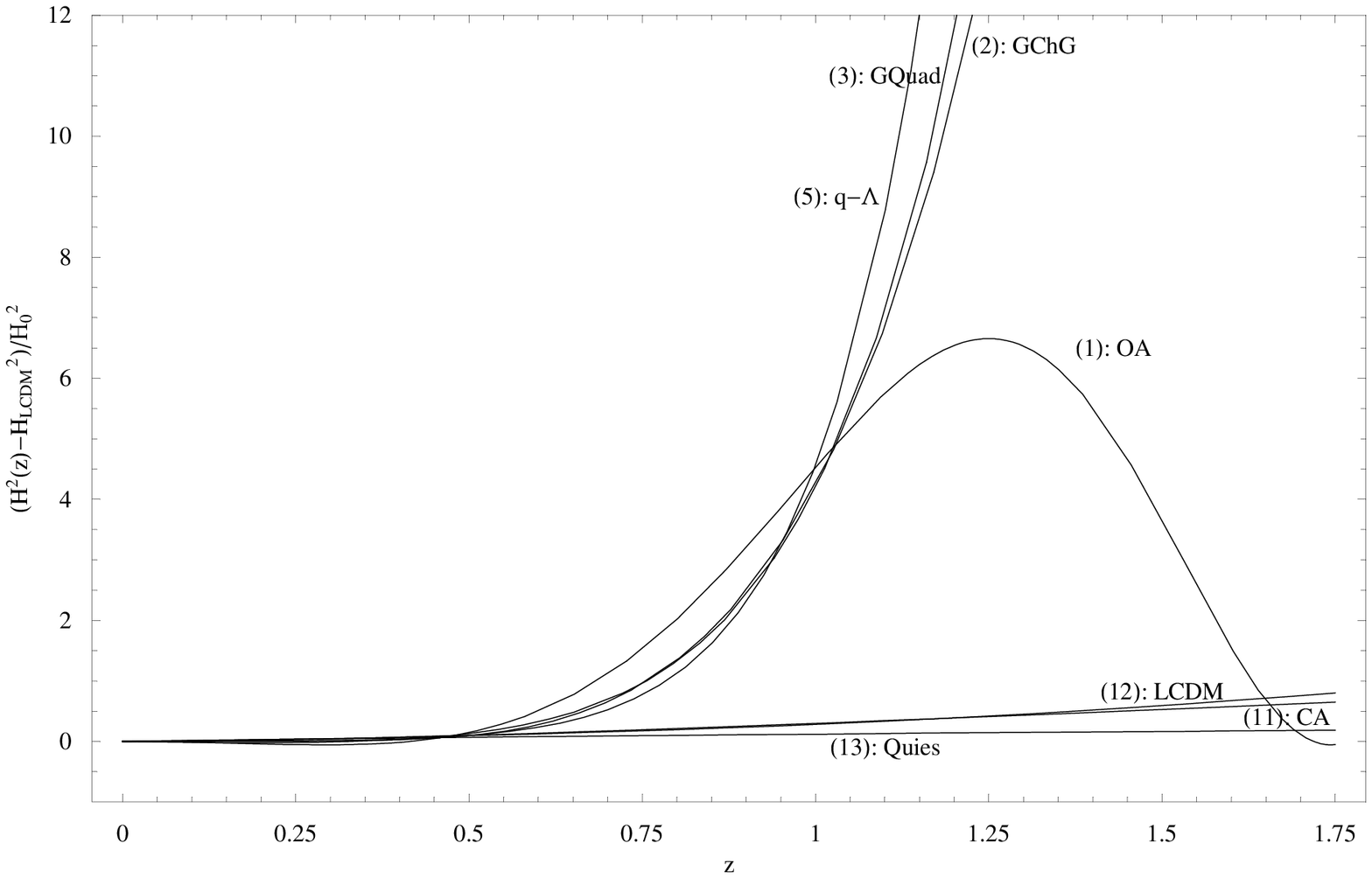}
\caption{The reduced Hubble parameter for some of the best and the
worst fits of the cosmological ansatze of Table 1. The number in
the parenthesis shows the rank $(1-13)$ of the corresponding
ansatz in terms of goodness of fit. The LCDM curve is not flat at
zero because in its construction we used the best fit value
$\Omega_{0m}=0.34$ while the $H_{LCDM}$ on the axis assumes the
prior $\Omega_{0m}=0.30$.} \label{fig4}
\end{figure}
From the theoretical viewpoint this idea is also supported for two
reasons:
\begin{itemize}
\item {\it Coincidence Problem Resolution:} An oscillating
expansion rate can help resolve the coincidence problem since our
present accelerating phase is viewed simply as part of a sequence
of accelerating and decelerating periods in the expansion history
of the universe \cite{Dodelson:2001fq,Griest:2002cu}

\item {\it Extra Dimensions:} Models with extra dimensions
generically predict oscillations of the stabilized modulus of the
extra dimension size (the radion field) due to its coupling to
redshifting matter
\cite{Perivolaropoulos:2002pn,Perivolaropoulos:2003we}. These
oscillations backreact on the expansion rate and induce
oscillations of the Hubble parameter.
\end{itemize}

Another factor pointing towards oscillating expansion rate is the
north-south pencil beam survey of Ref.\cite{Broadhurst:be} which
suggests an apparent periodicity in the galaxy distribution. The
number of galaxies as a function of redshift seems to clump at
regularly spaced intervals of $128h^{-1} Mpc$. Recent simulations
\cite{Yoshida:2001} have indicated that this regularity has a
priori probability  less than $10^{-3}$ in CDM universes with or
without a cosmological constant. An oscillating expansion rate
could resolve this puzzle without invoking special features in the
primordial fluctuations spectrum.

\section{Conclusion}

We have fitted several cosmological models using the maximum
likelihood method and the most recent SnIa data consisting of 194
data points. No priors have been used in our analysis other than
those indicated by other observations \ie flatness and
$\Omega_{0m} = 0.3$. 
The fact that we have fixed $\Omega_{0m}$ instead of marginalizing
over it could have artificially decreased somewhat the error bars
of the parameters. This decrease however is not important in view
of the main goal of this work which is the classification of the
models studied according to their quality of fit to the data.

We have confirmed recent studies
\cite{Alam:2003fg,Wang:2003gz,Wang:2004ru} indicating an increase
of the equation of state parameter with redshift termed
`metamorphosis' of dark energy in a recent
study\cite{Alam:2003fg}. We have shown however that the best fit
ansatze indicate that this `metamorphosis' continues beyond
$w(z)=0$ and leads to $w(z)>0$ at $z\simeq 1$. Nucleosynthesis and
large scale structure constraints are not consistent with $w(z)>0$
at arbitrarily high redshifts. Thus our best fits to the data can
only be made consistent with these constraints if some kind of
oscillating behavior is realized for the effective dark energy
equation of state parameter $w(z)$. This possibility is further
enhanced by the fact that an oscillating expansion rate ansatz has
provided the best fit to the data among all the 13 ansatze
considered and also by other theoretical and observational
arguments discussed in the previous section.

At low redshifts ($z<0.5$) all the fitted ansatze approach $w(z)
\simeq -1$ with $w(z)$ approximately constant. Most (but not all)
of the better fits predict $w(z)$ slightly less than $-1$ (up to
$-1.3$) for some redshift range within $[0,0.5]$ but not
necessarily at $z=0$. There are good fits however (like the
$quiessence-\Lambda $ ansatz) with $\chi_{min}^2$ below average
for which $w(z) >-1$ for all $z$ implying that phantom
energy\cite{Caldwell:1999ew,Caldwell:2003vq} ($w<-1$) is
consistent with the data but is not necessarily more probable than
dark energy ($w>-1$). Since there are good fits with $w(z) > -1$
and rapidly increasing $w(z)$ for $z>0.5$, we conclude that even
if the prior $w>-1$ were used, with the proper ansatz we would
still be able to see the rapid increase of $w(z)$ with $z$. This
does not agree with the conclusion of Ref. \cite{Alam:2003fg} that
it is the use of priors that would hide the increase of $w(z)$
with redshift. Here we have shown that the cosmological ansatz
selection can also play a crucial role in revealing or hiding the
true expansion history of the universe. Other comparisons of
particular ansatze with the SnIa data can also be found in the
literature\cite{Carturan:2002si,Gong:2003st,Gong:2004sa,Virey:2004pr,Gong:2004gh}.

Thus we are led to an important question: {\it Is there a
systematic way to use the SnIa data in constructing a relatively
simple cosmological ansatz that will give the best possible fit to
the data given the number of parameters? } Addressing this
important issue will be the subject of a subsequent paper.

{\bf Acknowledgements:} We thank U. Alam , V. Sahni and J. Tonry
for useful clarifications on the analysis of the SnIa data. This
work was supported by the European Research and Training Network
HPRN-CT-2000-00152. 

\section{Appendix}

Here we demonstrate that the marginalization over the zero point
magnitude $\bar M$ defined in equation (\ref{bm1}) would have
negligible effect ($O(1\%)$) on our results. Any model will
predict the theoretical value $D_L^{th}(z;a_1,...,a_n)$ with some
undetermined parameters $a_i$ (\eg $\Omega_m, \Omega_{\Lambda})$.
The best-fit model is obtained by minimizing the
quantity\cite{Choudhury:2003tj}: \be \chi^2({\bar
M'})=\sum_{i=1}^N{{(log_{10}D_L^{obs}(z_i)-0.2\bar M
'-log_{10}D_L^{th}(z_i))^2}\over {(\sigma_{log_{10}D_L (z_i)})^2 +
({{\partial log_{10} D_L (z_i)}\over {\partial z_i}}
\sigma_{z_i})^2}} \ee where $\bar M '=\bar M-\bar M_{obs}$ is a
free parameter representing the difference between the actual
$\bar M$, (see equation (\ref{bm1})), and it's assumed value $\bar
M_{obs}$ in the data.

Uniform marginalization over $\bar M'$ corresponds to indegrating
over $\bar M '$ and therefore working with a $\bar \chi^2$ defined
by: \be \bar\chi^2=-2ln(\int_{-\infty}^{+\infty}
e^{-\chi^2/2}d\bar M ') \ee which after some manipulation
gives:\be \bar\chi^2=-2ln(\int_{-\infty}^{+\infty} e^{-{1\over
2}C\bar M '^2+B\bar M '-{A\over2}}d\bar M ')\ee where \be
A=\sum_{i=1}^N {a_i^2\over \sigma_i^2}=\chi^2 ({\bar
M}'=0)\;,\;B=0.2\sum_{i=1}^N {a_i\over \sigma_i^2}\ee and \be
C=0.04\sum_{i=1}^N{1\over \sigma_i^2}\ee with \be a_i=log_{10}
D_L^{data}-log_{10}D_L^{th}\ee Thus, the ``marginalized" over
$\bar M'$ $\chi^2$ is: \be \label{bchi1} \bar\chi^2=\chi^2 (\bar
M'=0)-{B^2\over C}+ln({C\over2\pi})\simeq \chi^2 (\bar M'=0)\ee
because in the cases considered the last two terms on the RHS of
equation (\ref{bchi1}) are of $O(1)$ while the first is $O(10^2)$
and therefore dominates over the others. Thus the effects of the
marginalization are of order $1\%$ and can be neglected. This same
conclusion has also been reached in Ref. \cite{DiPietro:2002cz}.

\end{document}